\documentclass[aps,prl,amsmath,amssymb,reprint,superscriptaddress]{revtex4-2}
\usepackage{graphicx}% Include figure files
\usepackage{bm}% bold math
\usepackage{threeparttable}

\begin{document}

\title{Bright Second Harmonic Emission from Photonic Crystal Vertical Cavity}

\author{Lun Qu $^{\ddag}$}
\affiliation{The Key Laboratory of Weak-Light Nonlinear Photonics, Ministry of Education, School of Physics and TEDA Applied Physics Institute, Nankai University, Tianjin 300071, People’s Republic of China}
\email{These authors contributed equally to this work.}

\author{Zhidong Gu $^{\ddag}$}
\affiliation{The Key Laboratory of Weak-Light Nonlinear Photonics, Ministry of Education, School of Physics and TEDA Applied Physics Institute, Nankai University, Tianjin 300071, People’s Republic of China}
\email{These authors contributed equally to this work.}

\author{Chenyang Li}
\affiliation{The Key Laboratory of Weak-Light Nonlinear Photonics, Ministry of Education, School of Physics and TEDA Applied Physics Institute, Nankai University, Tianjin 300071, People’s Republic of China}

\author{Yuan Qin}
\affiliation{The Key Laboratory of Weak-Light Nonlinear Photonics, Ministry of Education, School of Physics and TEDA Applied Physics Institute, Nankai University, Tianjin 300071, People’s Republic of China}

\author{Yiting Zhang}
\affiliation{The Key Laboratory of Weak-Light Nonlinear Photonics, Ministry of Education, School of Physics and TEDA Applied Physics Institute, Nankai University, Tianjin 300071, People’s Republic of China}

\author{Di Zhang}
\affiliation{The Key Laboratory of Weak-Light Nonlinear Photonics, Ministry of Education, School of Physics and TEDA Applied Physics Institute, Nankai University, Tianjin 300071, People’s Republic of China}

\author{Jiaxian Zhao}
\affiliation{The Key Laboratory of Weak-Light Nonlinear Photonics, Ministry of Education, School of Physics and TEDA Applied Physics Institute, Nankai University, Tianjin 300071, People’s Republic of China}

\author{Qiang Liu}
\affiliation{The Key Laboratory of Weak-Light Nonlinear Photonics, Ministry of Education, School of Physics and TEDA Applied Physics Institute, Nankai University, Tianjin 300071, People’s Republic of China}

\author{Chunyan Jin}
\affiliation{The Key Laboratory of Weak-Light Nonlinear Photonics, Ministry of Education, School of Physics and TEDA Applied Physics Institute, Nankai University, Tianjin 300071, People’s Republic of China}

\author{Lishuan Wang}
\affiliation{Tianjin Key Laboratory of Optical Thin Film, Tianjin Jinhang Technical Physics Institute, Tianjin 300080, People’s Republic of China}

\author{Wei Wu}
\affiliation{The Key Laboratory of Weak-Light Nonlinear Photonics, Ministry of Education, School of Physics and TEDA Applied Physics Institute, Nankai University, Tianjin 300071, People’s Republic of China}

\author{Wei Cai}
\affiliation{The Key Laboratory of Weak-Light Nonlinear Photonics, Ministry of Education, School of Physics and TEDA Applied Physics Institute, Nankai University, Tianjin 300071, People’s Republic of China}

\author{Huasong Liu}
\affiliation{Tianjin Key Laboratory of Optical Thin Film, Tianjin Jinhang Technical Physics Institute, Tianjin 300080, People’s Republic of China}

\author{Mengxin Ren}
\email{ren\_mengxin@nankai.edu.cn}
\affiliation{The Key Laboratory of Weak-Light Nonlinear Photonics, Ministry of Education, School of Physics and TEDA Applied Physics Institute, Nankai University, Tianjin 300071, People’s Republic of China}
\affiliation{Collaborative Innovation Center of Extreme Optics, Shanxi University, Taiyuan, Shanxi 030006, People’s Republic of China}

\author{Jingjun Xu}
\email{jjxu@nankai.edu.cn}
\affiliation{The Key Laboratory of Weak-Light Nonlinear Photonics, Ministry of Education, School of Physics and TEDA Applied Physics Institute, Nankai University, Tianjin 300071, People’s Republic of China}

\begin{abstract}
We present a study on photonic vertical cavities consisting of nonlinear materials embedded in photonic crystals (PhCs) for resonantly enhancing second harmonic generation (SHG). Previous attempts at SHG in such structures have been limited to efficiencies of 10$^{-7}$ to 10$^{-5}$, but we demonstrate here a high SHG efficiency of 0.28\% by constructing a vertical cavity with a lithium niobate membrane placed between two PhCs, which exhibits high quality resonances. Our results open up new possibilities for compact laser frequency converters that could have a revolutionary impact on the fields of nonlinear optics and photonics.\\
keywords: Second harmonic generation, lithium niobate, photonic crystal, vertical cavity.
\end{abstract}

\maketitle

\section{Introduction}
Second harmonic generation (SHG) is a nonlinear optical process that converts two identical photons into one photon with doubled frequency \cite{franken1961generation}. It behaves as a cornerstone of modern photonics which has found a wide range of applications in laser technology, quantum optics, and optical microscopy. However, when developing ultracompact nonlinear photonic devices, we encountered the bottleneck problem of low SHG conversion efficiency at the nanoscale. Specifically, traditional strong pumping and phase matching techniques are difficult to apply to nanoscale photonic devices, and it is challenging to improve SHG efficiency at micro- and nanoscale due to the intrinsic small interaction volume and weak nonlinearities of materials. How to achieve high nonlinear efficiency at the nanoscale has become a critical challenge to address.

\begin{figure*}[htbp]
  \centering
  \includegraphics[width=140mm]{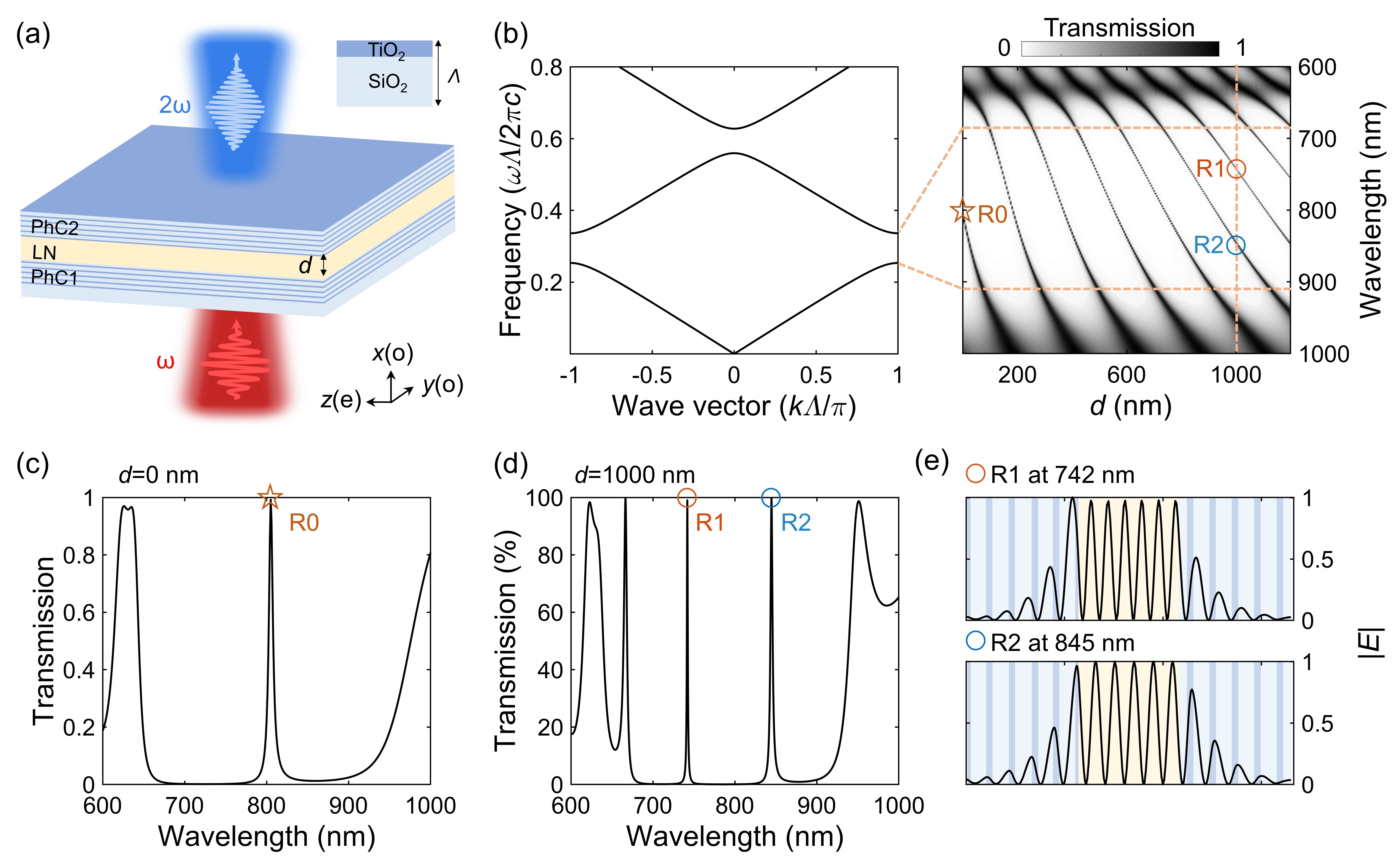}
  \caption{\textbf{Concept of photonic microcavity for SHG.} (a) The vertical cavity is formed by inserting an LN membrane (thickness of $d$) into two PhCs. Each PhC is made up of a stack of TiO$_2$ and SiO$_2$ alternating layers. A pump beam with frequency $\omega$ illuminates the structure, and the nonlinear signal at frequency 2$\omega$ is generated from the transmission direction. The unit cell of PhCs is detailed in the corner. The experimental coordinate was chosen to overlap with the LN principal crystallographic axes, where $z$-axis is defined along the optic axis ($e$) of LN. (b) Left panel: numerically calculated photonic band structure of the PhC. Two band gaps appear around the normalized frequencies of 0.3 and 0.6, respectively. Right panel: evolution of transmission spectra as a function of $d$. When $d$ increases from 0 to 1200~nm, the edge resonance R0 is pushed down to the lower band, and a sequence of discrete cavity modes are pulled down from the upper band into the band gap. (c) Simulated transmission spectrum of the structure with $d$=0, which corresponds to a direct contact of two PhCs without LN. A resonance peak (R0) associating to an edge state appears around 805~nm. (d) Insertion of a 1000~nm thick LN film results in two cavity modes at 742 (R1) and 845~nm (R2). (e) Electric field distribution inside the structure under $z$-polarized pump light at two resonances R1 and R2. Electric fields are strongly confined in the LN layer, and decay rapidly outside the LN layer.}
\label{fig1}
\end{figure*}

Notably, recent advances in nanotechnology have offered promising solutions to these challenges. There has been a broad consensus on the strategy to enhance the nonlinear response through optical resonance at nanoscale. The strong electric field confinement is the main reason for boosting the nonlinear efficiency when the incident light is close to the resonant frequency \cite{kuznetsov2016optically, koshelev2020subwavelength}. Recent years have witnessed remarkable successes in achieving high performance SHG using such as micro-rings or disks with ultrasharp resonances of linewidths below 0.01~nm \cite{guo2016second, liu2017cascading, lu2019periodically, lin2019broadband, chen2019ultra}. Although the light-matter interaction is significantly boosted in these resonators, they are highly dispersive, and are therefore primarily suitable for continuous wave lasers rather than pulsed light. Here we highlight the strategy for enhancing nonlinear response of pulsed lasers through optical resonances in photonic crystals (PhCs), which are artificial structures with spatially modulated dielectric functions, have manifested as a versatile photonic platform for engineering optical resonances \cite{joannopoulos2008molding}. Particularly, by stacking alternate layers of media with high and low refractive indices, one-dimensional (1D) PhCs can be constructed, which are favored for their advantages of simple design and easy fabrication. Furthermore, if a material slab is embedded inside PhCs as a defect layer, the perfect periodicity of the PhCs is broken, and resonant defect states are produced in the photonic band gap. The defect modes in this case are confined in the defect area by the surrounding PhC structure, forming a PhC vertical microcavity \cite{ren2009giant}. The emergence of defect states provides an efficient way to strongly localize photons, forming bright hotspots inside microcavity, which enhance nonlinear light-matter interactions. In the vertical cavities, both fundamental and SHG fields are propagating perpendicularly to the surface. Such cavities are believed compatible with a wide range of free-space applications, as well as technologies of other vertical devices, such as displays, optical data storage, optical steering, and so on \cite{jun2021full, zeng2015enhanced, hu2019coherent}. Attracted by their vertical emission feature and wide applications, the PhC vertical cavities for enhancing SHG have been extensively studied over the past decades. For example, the SHG from a chromophore monolayer embedded in the air spacer between two PhCs was reported \cite{trull1995second}. Resonant SHG in a poled polymer layer squeezed between a silver mirror and a dielectric PhC \cite{cao2000large}, and in porous silicon film sandwiched between two PhCs were also demonstrated \cite{golovan1999generation}. However, the weak nonlinear susceptibilities of these nonlinear materials limit the SHG magnitudes. In addition, semiconductors with large second-order nonlinearity, such as GaAs and AlGaAs have been used as cavity materials, which are grown on GaAs/AlAs PhC multilayers by molecular beam epitaxy \cite{nakagawa1995second}. However, the low refractive index contrast of GaAs/AlAs restricts the quality factors (Q) of the cavity, and these semiconductors strongly absorb the SHG wave in visible. These severely suppress the nonlinear efficiency. More recently, two-dimensional materials were placed inside PhC cavities to increase the likelihood of nonlinear response \cite{yi2016optomechanical,  day2016microcavity, liu2020nonlinear, zhao2022nonlinear}. So far, the SHG efficiency from the vertical microcavities is still limited to the order of 10$^{-7}$ to 10$^{-5}$, which is much far below the practical requirements.

In this paper, we highlight the strategy of enhancing nonlinear response through optical resonance, and discuss the use of PhCs vertical cavity embedded with a lithium niobate (LN) membrane as a promising platform to enhance the SHG conversion efficiency \cite{wang2017metasurface,jiang2018nonlinear,luo2018highly,lu2019periodically,li2022efficient,fedotova2020second,ma2021nonlinear,yuan2021,qu2022giant}. We note that LN is a particularly suitable material for this approach due to its high nonlinear susceptibility and excellent transparency over a wide spectral range \cite{weis1985}. In particular, the emerging lithium niobate on insulator (LNOI) provides one of the few ways to obtain the high quality single crystalline nonlinear optical membranes for nanophotonic devices \cite{xu2020high,sun2020microstructure,gao2021long,chen2022advances,fedotova2022lithium,boes2023lithium}. Our approach results in high-quality resonances that produce bright SHG spots observable by naked eyes. A record-high absolute conversion efficiency of 0.28\% is achieved, which is orders of magnitude higher than previous results from similar microcavities and other vertical emission SHG nanostructures. Our results pave a promising way to construct ultracompact laser frequency converters for photonics and optoelectronics.

\section{Results}
\subsection{1.Design and fabrication of photonic microcavity}

The vertical cavity comprises of two 1D PhCs (labeled as PhC1 and PhC2 in Figure 1a). Each PhC here is made up of two alternating dielectric slabs of titanium dioxide (TiO$_2$) and silicon dioxide (SiO$_2$). The thicknesses of these layers are $d_{\rm TiO_2}$=60~nm and $d_{\rm SiO_2}$=170~nm, respectively. The unit cells of the two PhCs are indicated in the upper right corner, which have a lattice period of $\Lambda$=$d_{\rm TiO_2}$+$d_{\rm SiO_2}$=230~nm. The band structures of the two PhCs are calculated in Figure 1b, where the refractive indices of TiO$_2$ and SiO$_2$ were set as $n_{\rm TiO_2}$ = 2.36 and $n_{\rm SiO_2}$ = 1.46 in the calculation, respectively. Two band gaps appear around the normalized frequencies of 0.3 and 0.6, respectively. The geometric parameters of the PhCs are properly designed to ensure that the lowest band gap covering the range from 700 to 900~nm, which is accessible by a typical Ti-sapphire femtosecond (fs) laser. An edge state presents within this lowest band gap if the two PhCs are put together directly. This edge state demonstrates itself as a resonance peak (R0) in transmission spectrum in the band gap, as marked by a star in Figure 1c.

\begin{figure}[htbp]
  \centering
  \includegraphics[width=85mm]{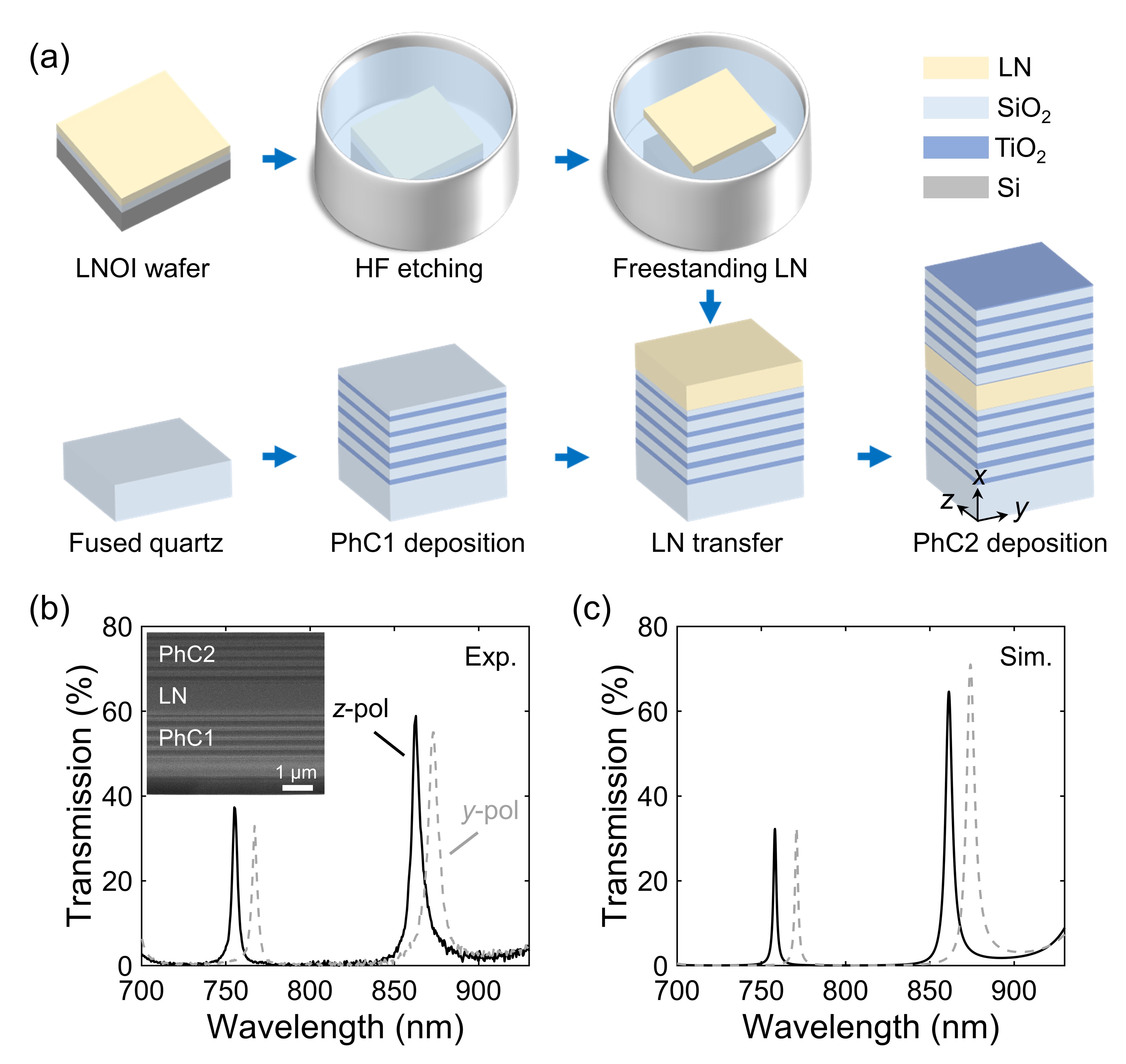}
  \caption{\textbf{Microcavity fabrication and linear spectral characterization.} (a) Main steps for fabricating the microcavity. An $x$-cut LN film is first detached from an LNOI wafer, and is then sandwiched between two PhCs on a fused quartz substrate. (b) Measured transmission spectra of the microcavity under normal incident $z$- (solid line) and $y$-polarized light (dashed line). The spectra show a distinct spectral shift, due to the birefringence of LN. A cross-section scanning electron microscopy (SEM) image of the microcavity is placed in the inset. (c) Simulated transmission spectra considering a non-zero imaginary part $\kappa$ of 0.0018 in the refractive index of LN.}
  \label{fig2}
\end{figure}

To enable the SHG, a nonlinear crystalline LN film of thickness $d$ is further sandwiched in the interface. To utilize the largest second-order susceptibility ($d_{33}$) of the LN, the LN film is $x$-cut, and its optic axis ($e$) locates within the LN plane. As shown in the right panel of Figure 1b, when $d$ increases, the mode R0 has more space to oscillate, thus its frequency decreases. As $d$ becomes larger than 150~nm and further increases, the resonance R0 is pushed down to the lower band, and a series of discrete cavity modes are pulled down from the upper band into the band gap. The designed LN film has a thickness of approximately 1000~nm, which closely matches the nonlinear coherence length of LN in the spectral range of 700 to 900~nm. This benefits maximizing the SHG efficiency in situations where phase mismatching occurs \cite{ma2020second}. Two resonances appear around 742 and 845~nm, which are labeled by R1 and R2 in Figure 1d, respectively. In addition, as confirmed by numerical simulations in Figure 1e, the electric fields are tightly confined within the LN film at two resonances, and are evanescent rapidly outside the LN layer. Resulting from such field localization at the resonances, the enhanced SHG is expected.

Figure 2a presents the fabrication flowchart of the vertical cavity. An $x$-cut LN membrane was first detached from an LNOI wafer (NANOLN Co. Ltd.) by wet etching, which was then transferred to the interface between two PhCs on a fused quartz substrate (fabrication details are given in Methods). A cross-section scanning electron microscopy (SEM) image of the fabricated sample is inserted in Figure 2b. To facilitate sample fabrication, both PhC1 and PhC2 consist of five periods in our experiment. We measured the transmission spectra using a commercial microscopic-spectrometer (IdeaOptics Co. Ltd.), as shown in Figure 2b. The solid line gives the spectrum excited by $z$-polarized light, which corresponds to extraordinary ray incidence. Two transmission peaks appear at 754 and 861~nm. By fitting the experimental spectra with a Fano curve, and dividing the central wavelength by the resonance width, the Q-factor of resonances R1 and R2 are yielded as 200 and 144, respectively. Furthermore, since LN is an anisotropic crystal, the spectrum experiences a significant redshift when the incident light changes to $y$-polarized (dashed line). The deviation of the transmittance from 100\% implies the existence of optical losses in our sample. Such optical losses are possibly introduced by inevitable chemical contamination during the transfer of LN film. To investigate the influence of the optical losses on the resonances, we introduced a non-zero imaginary part $\kappa$ of 0.0018 to the refractive index of LN in simulation, and the calculated results (Figure 2c) agree with the experimental ones (Figure 2b) very well.

\subsection{2. Bright SHG from photonic microcavity}
\begin{figure}[htbp]
  \centering
  \includegraphics[width=85mm]{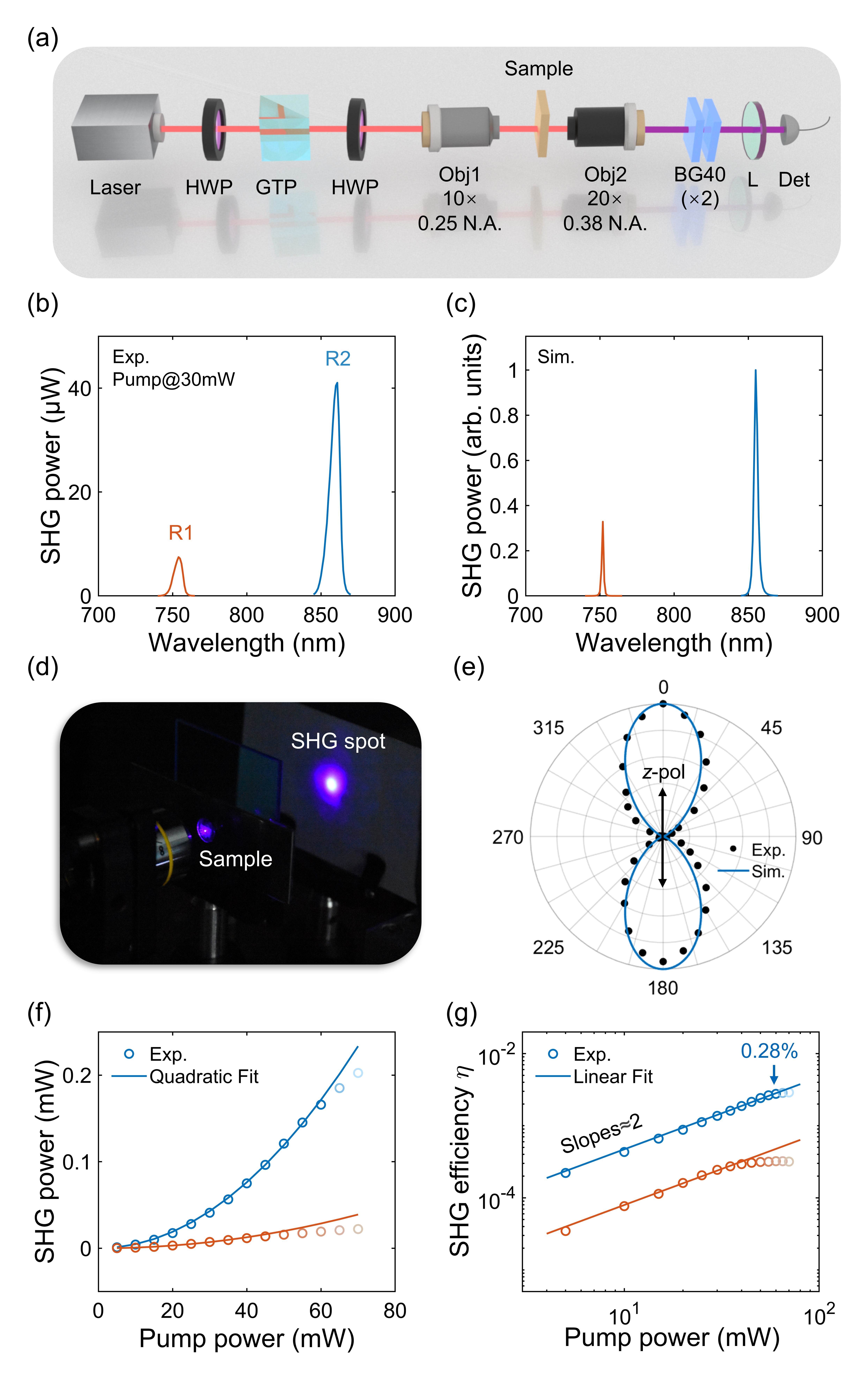}
  \caption{\textbf{Experimental observation of bright SHG from microcavity.} (a) A schematic illustration of our experimental setup. HWP: half-wave plate, GTP: Glan-Taylor prism, Obj: objective, BG40: band-pass filter, L: lens, Det: power meter. (b) The measured SHG power for pump wavelength scanning near resonances R1 (orange curve) and R2 (blue curve). The pump power was set at 30~mW during measurements. (c) The simulated SHG power for pump wavelength scanning near resonances R1 (orange curve) and R2 (blue curve). (d) Photograph of a bright SHG spot observed in experiments without diffraction. (e) Polar plots of the SHG intensity as a function of the pump polarization azimuth at resonance R2. The black dots represent the experimental data while the solid line lobes represent the simulated SHG strength as a function of the input polarization. 0$^{\circ}$ and 90$^{\circ}$ correspond to $z$- and $y$-polarization, respectively. (f) Measured SHG power under different pump power. Circles represent the measured points, and solid lines are quadratic eye-guides. The data points deviate from the quadratic relationship at higher pump power (light colored data points) due to the saturation effect. (g) SHG conversion efficiency $\eta$ as a function of the pump power in a log-log scale coordinate, and solid lines are linear eye-guides.}
\label{fig3}
\end{figure}

To experimentally characterize the resonance enhanced SHG properties, the sample was illuminated by a tunable Ti:sapphire femtosecond laser (Maitai, Spectra-Physics), as shown in Figure 3a. The $z$-polarized pump laser beam was focused onto the sample by an objective, and the generated SHG signals were measured using a power meter (S130VC, Thorlabs). The introduction to the optical setup is detailed in Methods. To demonstrate the relationship between the SHG enhancements and the resonances, we measured the SHG power by sweeping the pump wavelength around two resonances, i.e., from 740 to 765~nm, and from 845 to 870~nm with a step of 1~nm. While adjusting the pump wavelength, the average pump power was kept constant at 30~mW. As shown in Figure 3c, strong SHG peaks are observed at two resonances. Furthermore, the SHG from resonance R2 is more than 4 times stronger than R1. We further simulated the SHG power near R1 and R2 (Figure 3c), which presents good agreement with the experimental results. A bright bluish-purple SHG spot was seen by naked eyes (as shown in the photograph in Figure 3d) in the transmission direction under 861~nm pumping (corresponding to R2), confirming the excellent SHG performance of our sample. Furthermore, compared with metasurfaces with lateral structural patterns that usually diffact the SHG \cite{fedotova2020second,qu2022giant}, our microcavity presents a feature of non-diffrcation in the SHG, which benefits the real applications. Moreover, the SHG is expected to highly depend on the pump light polarizations. Because, first, the $d_{33}$ is about 4.5 times larger than the other nonlinear tensor elements, resulting in stronger SH light by the $z$-polarized pump light \cite{ma2020second}; second, birefringence induced polarization dependent resonances in Figure 2b also lead to different field localizations and different nonlinear responses at a certain pump wavelength. As a result of changes in the polarization of the pump light, an ideal 8-like pattern is achieved, with maxima located at 0$^\circ$ and 180$^\circ$, as shown in Figure 3e of the result of R2. In accordance with the second order nature of this process, the generated SHG power presents a quadratic dependence on the average input laser power before saturation, as shown in Figure 3f. The SHG power at R2 achieves 0.166~mW as the pump power reaches 60~mW (corresponding to the peak intensity of about 6.3 GW/cm$^2$), which is more than 10-times stronger than R1 before saturation (0.014 mW SHG under 45~mW pump). Notably, an absolute SHG conversion efficiency ($\eta$, calculated as a ratio of SHG average power to that of the pump light) of 0.28\% is attained at R2 without reaching saturation from Figure 3g. As shown in Figure 4, this efficiency is at least two orders of magnitude higher than the values of previous PhC vertical microcavities on the level of 10$^{-7}$ to 10$^{-5}$ \cite{nakagawa1995second, yamada1996continuous, simonneau1997second}, and also higher than that of vertical emission all-dielectric metasurfaces on the level of 10$^{-6}$ to 10$^{-4}$ \cite{fedotova2020second, ma2021nonlinear, yuan2021, qu2022giant, carletti2021steering, huang2022resonant, liu2016resonantly, vabishchevich2018enhanced, anthur2020continuous}. Noteworthily, benefiting from the high damage threshold of the LN, our microcavity present the largest SHG average power, which is important for practical applications.

\begin{figure}[htbp]
  \centering
  \includegraphics[width=70mm]{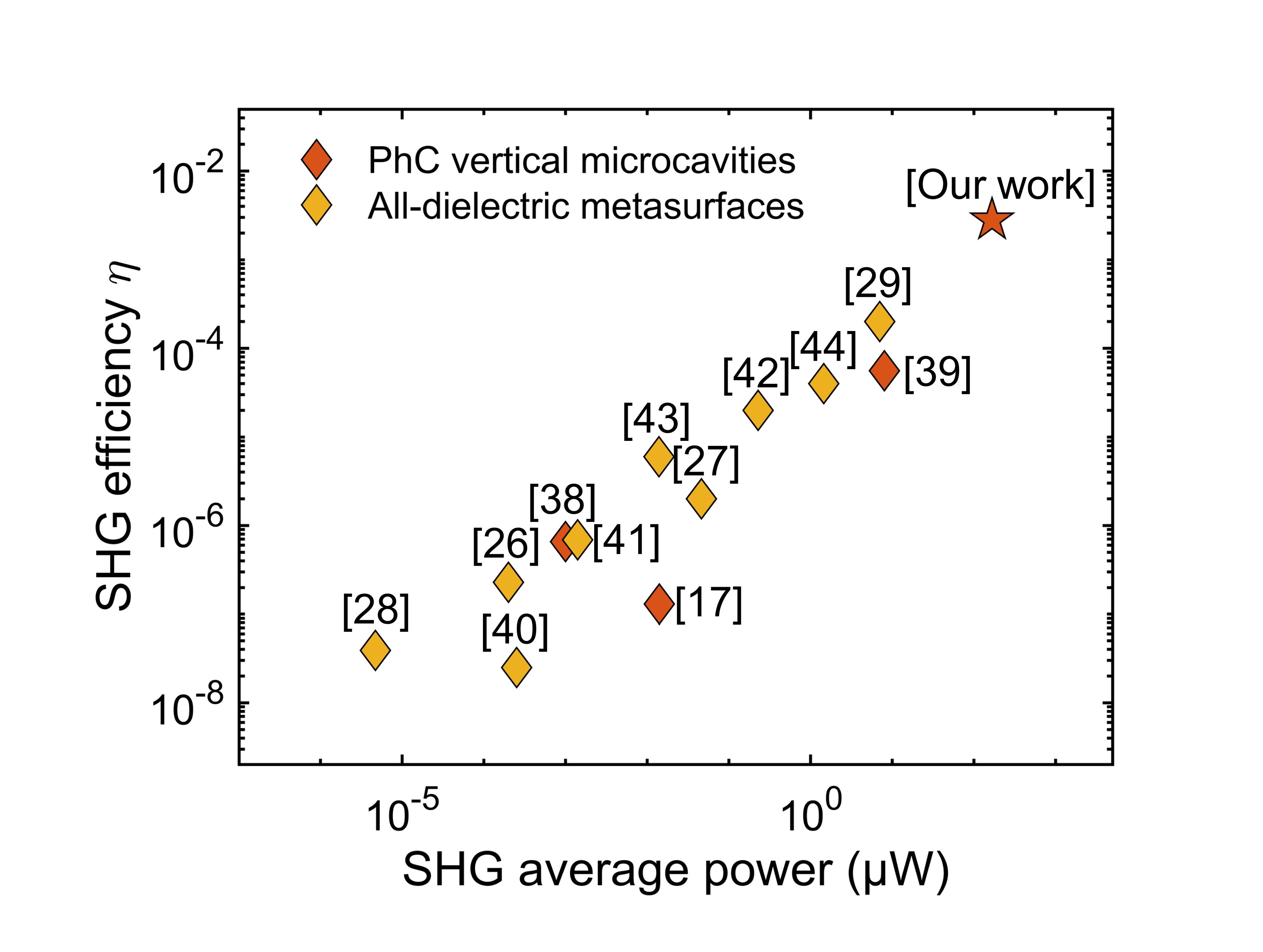}
  \caption{\textbf{Comparison of SHG conversion efficiency from different structures.} SHG efficiency $\eta$ and SHG average power are shown on the vertical and horizontal axes, respectively.
  }
\label{fig4}
\end{figure}

\section{Discussions}
In conclusion, we have experimentally observed bright SHG emission from a vertical cavity, which is formed by inserting an LN film between two PhCs. By pumping at the resonance wavelengths, a high SHG efficiency of 0.28\% was achieved, which is 2 orders of magnitude larger than previous similar microcavities. Although only SHG was discussed here, our results can be applied to study other phenomena such as sum-frequency generation, third harmonic generation, four-wave mixing, optical parametric amplification, and so on. Furthermore, within the prospective of enhanced SHG efficiency, its inverse process referred as spontaneous parametric down conversion is also accessible, which is essential for entangled photon pairs generation. We believe our results provide a promising platform to develop ultracompact efficient nonlinear light sources for both classical and quantum photonic applications.

\section*{Methods}

\noindent\textbf{Sample fabrication.} The process started with an LNOI wafer (1$\times$1~cm$^2$ in area), which consists of a 1030 nm thick single crystalline LN thin film bonded to a silicon (Si) substrate via a 2 $\mu$m thick SiO$_2$ buffer layer. The wafer was immersed in hydrofluoric (HF) acid solution to etch the SiO$_2$ layer. Consequently, the LN thin film detached from the Si substrate and floated in the solution. The PhC1 was deposited on a fused quartz substrate by ion beam sputtering deposition technique, onto which we transferred the LN membrane. After further depositing the PhC2 on top, the sample was fabricated.\\

\noindent\textbf{Nonlinear measurement.} The sample was illuminated by a tunable Ti:sapphire femtosecond laser (Maitai, Spectra-Physics). The pump laser beam was focused onto the sample by a 10$\times$ objective (N.A.=0.25) forming a spot with a diameter of about 10~$\mu$m. The pump intensity was adjusted by the first half wave plate (HWP) before the Glan–Taylor prism (GTP), while the pump polarization was adjusted by the second HWP behind. The SH signal was collected by a 20$\times$ objective (N.A.=0.38), and the transmitted pump wave was filtered out by BG40 colored glasses. The SH power was measured using a silicon photodiode power meter (S130VC, Thorlabs) and calibrated by considering the transmittance of the collection objective and BG40 glasses.\\

\medskip
\noindent \textbf{Acknowledgements} \par
This work was supported by National Key R\&D Program of China (2022YFA1404800, 2019YFA0705000); Guangdong Major Project of Basic and Applied Basic Research (2020B0301030009); National Natural Science Foundation of China (12222408, 92050114, 12174202, 12074200); China Postdoctoral Science Foundation (2022M721719, 2022M711710); 111 Project (B23045); PCSIRT (IRT0149); Fundamental Research Funds for the Central Universities. We thank Nanofabrication Platform of Nankai University for fabricating samples.

\medskip
\noindent \textbf{Author Contributions} \par
 L.Q. and Z.G. contributed equally to this work. L.Q., Z.G. and M.R. conceived and performed the design. L.Q., C.L., Y.Q., Y.Z., D.Z., J.Z. and W.C. performed numerical simulations and analyzed data. Z.G., C.J., L.W., W.W. and H.L. fabricated samples. L.Q., D.Z. and Q.L. performed optical measurements. M.R. and J.X. organized and led the project. All authors discussed the results and prepared the manuscript.

\medskip
\noindent\textbf{Data availability}\par
All relevant data supporting the results of this study are available within the article and its supplementary information files. Further data are available from the corresponding authors upon request.

\medskip
\noindent \textbf{Competing interests} \par
The authors declare no conflict of interest.

\bibliography{bibliography}

\end{document}